\documentclass[a4paper,11pt]{article}
\usepackage{pos}

\newcommand{\Mp}{M_{\rm Pl}}
\newcommand{\M}{M}
\newcommand{\R}{R_{\textrm{fs}}}
\newcommand{\F}{Q_s}

\title{Inflation with strongly non-geodesic motion: theoretical motivations and observational imprints}
\ShortTitle{Inflation with strongly non-geodesic motion}

\author*[a]{S\'ebastien Renaux-Petel}

\affiliation[a]{Institut d'Astrophysique de Paris, GReCO, UMR 7095 du CNRS et de Sorbonne Universit\'e,\\
98bis boulevard Arago, 75014 Paris, France}

\emailAdd{renaux@iap.fr}

\abstract{A new class of inflationary attractors characterized by a strongly non-geodesic motion has been discovered and explored in the past few years. I describe how they naturally arise in negatively curved field space, allowing to inflate on potentials that are steep in Planck units, albeit without alleviating the ever-present naturalness issue of inflation. In these scenarios, fluctuations often experience a transient tachyonic instability, which can be described by a single-field effective field theory with an imaginary speed of sound. Independently of the precise ultraviolet origin of the latter, this leaves a peculiar imprint in the form of a high-level of primordial non-Gaussianities of flattened type for all higher-order correlation functions. On small scales, a transient phase of strongly non-geodesic motion provides a mechanism to generate primordial black holes and can leave specific signatures in the form of oscillations in the frequency profile of the stochastic gravitational wave background.}

\FullConference{%
  *** The European Physical Society Conference on High Energy Physics (EPS-HEP2021), ***\\
  *** 26-30 July 2021 ***\\
  *** Online conference, jointly organized by Universität Hamburg and the research center DESY ***
}

\begin{document}
\maketitle

Inflation acts as giant microscope: it takes a tiny patch of space and stretches it to become the entire observable universe. In this process, vacuum quantum fluctuations are also stretched to cosmological scales, and provide the seeds of all structures in the universe. Inflation has passed very stringent tests \cite{Akrami:2018odb,Akrami:2019izv} and has become the paradigm of the primordial universe, but its physics is still elusive. In this respect, one of its key feature is its ultraviolet sensitivity, which manifests, in the simplest realization of inflation, in the form of the so-called $\eta$ problem. Namely, in order to generate a prolonged phase of expansion with almost constant Hubble rate $H$ with a single scalar field rolling down its potential, the latter should be flat compared to the Planck scale, with $\epsilon \equiv -\dot{H}/H^2 \simeq \frac12\Mp^2 (V'/V)^2 \ll 1$ and $\eta=\Mp^2 V^{''}/V \ll 1$, corresponding to an unnaturally small inflaton mass $\eta \sim m_\phi^2/H^2 \ll 1$. However, even Planck-suppressed operators in general spoil the flatness of the potential.
This sensitivity of inflation to Planck-scale physics motivates its study in string theory, which makes explicit the challenge of embedding inflation in high-energy physics, typical situations including many degrees of freedom with steep potentials and large couplings \cite{Baumann:2014nda}. This has prompted the discovery of inflationary mechanisms different from slow-roll, and showed that the latter can be considered at best as an emergent approximate description.

In this context, a typical framework that encompasses large classes of top-down constructions is the one of nonlinear sigma-models, of Lagrangian ${\cal L}=-\frac12 G_{IJ}(\boldsymbol{\phi}) \partial^\mu \phi^I \partial_\mu \phi^J-V(\boldsymbol{\phi})$. Here, the fields $\phi^I$ interact through potential
and kinetic interactions, and it is useful to adopt a geometrical picture in which the $\phi^I$s are considered as coordinates in an abstract ``field space'', whose metric is $G_{IJ}$, and to formulate the physics in a form that is manifestly covariant under field redefinitions. An important feature is that this field space is generically curved, with a non-zero curvature (tensor) $\R \sim 1/\M^2$ characterized by an energy scale $\M$. Such Lagrangians emerge as low-energy 4D effective descriptions of string theory, hence with a cutoff lower than the Kaluza-Klein and string scales, themselves lower than the Planck mass, so that a large hierarchy $\M/\Mp \ll 1$ is built-in. However, this hierarchy makes dangerous some derivative interactions that would be otherwise harmless. Considering interactions between the would-be inflaton $\phi$ and another scalar field $\chi$, corrections to the kinetic term of the type ${\cal L} \supset c  (\partial \phi)^2 \frac{\chi^2}{\M^2}$ give a contribution to the mass of $\chi$ of order
$\frac{\Delta m_{\chi}^2}{H^2} \sim c \frac{\dot \phi^2}{ H^2 \M^2} \sim c\, \epsilon  \left(\frac{\Mp}{\M}\right)^2$. The large hierarchy of scales can easily compensate the $\epsilon$ suppression, making this contribution comparable to or larger than the bare mass $m^2/H^2$ coming from the potential, completely altering the mass spectrum of the theory. 
This has important consequences beyond this example \cite{Renaux-Petel:2015mga}:
the simple fact that initially neighboring geodesics tend to fall away from each other in a negatively curved field space implies that the rolling of the inflaton there tends to induce a destabilization of inflationary trajectories.

When this \textit{geometrical destabilization} takes place, the system bifurcates into another attractor inflationary solution \cite{Grocholski:2019mot}, first studied in \cite{Renaux-Petel:2015mga} and in more detail in \cite{Garcia-Saenz:2018ifx}, where it was called \textit{sidetracked inflation} owing to its characteristics. 
To better appreciate them, consider the fields' dynamics:
${\cal D}_t \dot \phi^I+3 H \dot \phi^I+ V^{,I}=0$
with ${\cal D}_t A^I \equiv \dot{A^I} + \Gamma^I_{JK} \dot \phi^J A^K$ the covariant derivative along the trajectory. The vanilla multifield slow-roll dynamics corresponds to fields following the gradient flow of the potential, $\dot \phi^I \simeq - V^{,I}/3H$, hence with ${\cal D}_t \dot \phi^I \simeq 0$ and a trajectory that is approximately a field space geodesic. However, a prolonged phase of inflation only requires a small covariant acceleration \textit{along} the inflationary trajectory: $e_{\sigma I} {\cal D}_t \dot \phi^I \simeq 0$, where $e_{\sigma I}=\dot \phi^I/|\dot \phi^I|$ is the unit vector tangent to the trajectory. Considering two fields for simplicity, the component of the acceleration pointing in the perpendicular direction $e_s^I$ is not constrained, i.e. the dimensionless perpendicular acceleration $\eta_\perp \equiv H^{-1} e_{sI} {\cal D}_t e_\sigma^I =-\frac{e_{s I} V^{,I}}{H |\dot \phi^I|}$ need not be small. This is what takes place in sidetracked inflation: a heavy field $\chi$ is displaced from its minimum and adiabatically follows the minimum of its \textit{effective} potential, which depends on the kinetic energy of the inflaton, like in the gelaton scenario \cite{Tolley:2009fg}. This can be heuristically understood as arising from the
balance between the repulsive force of the negatively curved field space and the stabilizing force of the potential, and as a result the inflationary trajectory strongly deviates from a field space geodesic, with $\eta_\perp \gg 1$. Moreover, the coupling between the two fields translates into an effective flattening of the potential for the inflaton (see also \cite{Dong:2010in,McAllister:2014mpa}).
As a result, this mechanism allows inflation on potentials that would otherwise be too steep for slow-roll inflation, with the weaker requirement that the potential be flat with respect, not to $\Mp$, but to the curvature scale $\M$ of the field space.
Actually, one can show kinematically that one can inflate on Planck-steep potentials
only at the condition that the motion is strongly non-geodesic \cite{Achucarro:2018vey}. The fact that this can naturally arise dynamically in a negatively curved field space is particularly interesting, and the study of such kind of attractors has been pursued actively in recent years (see for instance \cite{Cremonini:2010ua,Renaux-Petel:2015mga,Brown:2017osf,Mizuno:2017idt,Achucarro:2017ing,Christodoulidis:2018qdw,Linde:2018hmx,Garcia-Saenz:2018ifx,Garcia-Saenz:2018vqf,Achucarro:2018vey,Christodoulidis:2019mkj,Bjorkmo:2019aev,Fumagalli:2019noh,Bjorkmo:2019fls,Christodoulidis:2019jsx,Bjorkmo:2019qno,Aragam:2019khr,Bravo:2019xdo,Chakraborty:2019dfh,Aragam:2020uqi}). However, as already anticipated in \cite{Garcia-Saenz:2018ifx}, it is important to stress that inflation with strongly non-geodesic motion does not alleviate the difficulties to embed inflation in high-energy physics, which simply manifest themselves in a different manner than in slow-roll. In this respect, the comparison with DBI inflation 
\cite{Alishahiha:2004eh}, with Lagrangian ${\cal L}=-M^4 \left(\sqrt{1- \frac{\dot{\phi}^2}{M^4}}-1 \right)-V$, is interesting. In a strongly warped throat, corresponding to $\M \ll \Mp$, DBI inflation can also support inflation on Planck-steep potentials under the conditions
$\frac{M^2}{H \Mp}   \lesssim \frac{\Mp V'}{V} \ll  \frac{H \Mp}{M^2}$,    
while in a strongly curved field space, sidetracked inflation with a strongly non-geodesic motion requires $\frac{m M}{ H\Mp}   \lesssim \frac{\Mp V'}{V} \ll  \frac{ H\Mp}{m M}$. In each case, the first inequality states that the potential should be steep enough so that the inflaton rolls sufficiently fast, either to approach the speed limit or to be geometrically destabilized. The second inequalities state the more stringent conditions to actually support a prolonged phase of inflation in a manner very different from slow-roll, either in the relativistic regime of DBI inflation, or in the sidetracked manner. However, in each case, in a theory with a cutoff $M \ll \Mp$, the natural expectation is that the potential varies substantially over a distance $M$ in field space, i.e.~$M V'/V \sim 1$, so that the conditions above are equally technically unnatural than requiring a flat potential with respect to $\Mp$ in canonical slow-roll inflation (see e.g.~\cite{Chen:2008hz,Baumann:2014nda} for discussions of the tuning required in DBI inflation).

Despite this ever-present tuning of inflationary scenarios, slow-roll, DBI or sidetracked-type inflation are all interesting genuinely different mechanisms, which are important to put to the test through their predictions for primordial density fluctuations. As we now explain, in the same way as large equilateral non-Gaussianity would indicate a low speed of sound of inflationary fluctuations, with DBI inflation as a chief example, large non-Gaussianities of flattened type would be a smoking-gun signature of an imaginary sound speed, with inflation with strongly non-geodesic motion as a prime example \cite{Garcia-Saenz:2018ifx,Garcia-Saenz:2018vqf,Fumagalli:2019noh,Bjorkmo:2019qno,Ferreira:2020qkf}. For this, note that in nonlinear sigma models, the quadratic Lagrangian for fluctuations read ${\cal L}^{(2)}=a^3\bigg[\Mp^2\epsilon\left(\dot{\zeta}^2-\frac{(\partial \zeta)^2}{a^2}\, \right)+2 | \dot \phi^I|\eta_{\perp}\dot{\zeta}\F+\frac{1}{2}\left(\dot{Q}_s^2-\frac{(\partial \F)^2}{a^2}-m_s^2\F^2\right) \bigg]$ with $\zeta$ the comoving curvature perturbation and $Q_s$ the entropic fluctuation of mass $m_s^2/H^2=V_{;ss}/H^2-\eta_\perp^2+ \epsilon \R \Mp^2$ (see \cite{Garcia-Saenz:2019njm,Pinol:2020kvw} for the cubic Lagrangian). A strongly non-geodesic motion, characterized by a large ``bending'' parameter $\eta_\perp$, automatically comes with a large coupling between the two fluctuations --- one that cannot be treated perturbatively --- and it also results in a large negative contribution to the entropic mass --- which can be seen as a manifestation of the familiar centrifugal force ---, something that is only reinforced by the negative geometrical contribution to the mass in a negatively curved field space. Without a stabilization from the potential, it is thus a built-in feature of these scenarios that fluctuations experience a tachyonic instability after momenta $k/a$ drop below $|m_s|$. This instability is only transient though, as a large bending instead contributes positively to the effective super-Hubble entropic mass, signaling the stability of the background. 
The resulting dynamics of fluctuations can be described by the single-field effective field theory of inflationary fluctuations, albeit in the unconventional situation of an imaginary sound speed $c_s^2<0$. This largely unexplored regime is worth studying by itself, with inflation with strongly non-geodesic motion providing a (partial) UV-completion of it. Indeed, this description is naturally valid at low energy only, when modes satisfy $k |c_s| /a < x H$, with $x \sim \eta_\perp \gg 1$ in the situation of interest. Due to the exponential growth of fluctuations, the power spectrum is largely amplified by $\sim e^{2x}$ compared to standard scenarios, but this is not observable by itself (this does result in typically larger deviations to scale-invariance than standard, although this can be tuned). However, this leaves a striking signature in non-Gaussianities, with an enhancement of the bispectrum near flattened triangular configurations 
$f_{\textrm{NL}}^{\textrm{flat}} \sim  \left( \frac{1}{|c_s|^2}+1\right) x^3$ \cite{Garcia-Saenz:2018vqf}, due to constructive interferences between exponentially growing and decaying
modes, akin to what arises with excited initial states, albeit without oscillatory behaviour. The same phenomenon is at play for all higher-order correlation functions, which display a hierarchical enhancement for particular flattened shapes, guaranteeing perturbative control as soon as $f_{\textrm{NL}}^{\textrm{flat}} {\cal P}_\zeta^{1/2} \lesssim 1$ \cite{Fumagalli:2019noh,Bjorkmo:2019qno}. Note that a precise agreement has been found between this EFT treatment and numerical computations in concrete models with strongly non-geodesic motion that display this transient tachyonic instability \cite{Fumagalli:2019noh}. These results showing the high level of non-Gaussianities inherent to these models result in powerful model-independent constraints if these scenarios are to describe the phase of inflation probed by the CMB. On smaller scales, a transient phase of strongly non-geodesic motion provides a natural mechanism to boost the amplitude of primordial fluctuations and hence generate primordial black holes \cite{Palma:2020ejf,Fumagalli:2020adf}. Independently, if this phase is short enough, the primordial power spectrum exhibits order one oscillations characteristic of sharp features, that are transferred to oscillations in the frequency profile of the scalar-induced stochastic gravitational wave background \cite{Fumagalli:2020nvq}.

\acknowledgments I am grateful to my collaborators on this topic: Jacopo Fumagalli, Sebastian Garcia-Saenz, Lucas Pinol, John Ronayne, Krzysztof Turzy\'nski, Vincent Vennin and Lukas T. Witkowski. I am supported by the European Research Council under the European Union's Horizon 2020 research and innovation programme (grant agreement No 758792, project GEODESI).

\bibliographystyle{JHEP}
\bibliography{Biblio-2020}

\providecommand{\href}[2]{#2}\begingroup\raggedright\begin{thebibliography}{10}

\bibitem{Akrami:2018odb}
{\scshape Planck} collaboration, \emph{{Planck 2018 results. X. Constraints on
  inflation}}, \href{https://doi.org/10.1051/0004-6361/201833887}{\emph{Astron.
  Astrophys.} {\bfseries 641} (2020) A10}
  [\href{https://arxiv.org/abs/1807.06211}{{\ttfamily 1807.06211}}].

\bibitem{Akrami:2019izv}
{\scshape Planck} collaboration, \emph{{Planck 2018 results. IX. Constraints on
  primordial non-Gaussianity}},
  \href{https://arxiv.org/abs/1905.05697}{{\ttfamily 1905.05697}}.

\bibitem{Baumann:2014nda}
D.~Baumann and L.~McAllister, \emph{{Inflation and String Theory}}, Cambridge
  University Press (2015), [\href{https://arxiv.org/abs/1404.2601}{{\ttfamily
  1404.2601}}].

\bibitem{Renaux-Petel:2015mga}
S.~Renaux-Petel and K.~Turzy{\'n}ski, \emph{{Geometrical Destabilization of
  Inflation}},
  \href{https://doi.org/10.1103/PhysRevLett.117.141301}{\emph{Phys. Rev. Lett.}
  {\bfseries 117} (2016) 141301}
  [\href{https://arxiv.org/abs/1510.01281}{{\ttfamily 1510.01281}}].

\bibitem{Grocholski:2019mot}
O.~Grocholski, M.~Kalinowski, M.~Kolanowski, S.~Renaux-Petel, K.~Turzy{\'n}ski
  and V.~Vennin, \emph{{On backreaction effects in geometrical destabilisation
  of inflation}},
  \href{https://doi.org/10.1088/1475-7516/2019/05/008}{\emph{JCAP} {\bfseries
  1905} (2019) 008} [\href{https://arxiv.org/abs/1901.10468}{{\ttfamily
  1901.10468}}].

\bibitem{Garcia-Saenz:2018ifx}
S.~Garcia-Saenz, S.~Renaux-Petel and J.~Ronayne, \emph{{Primordial fluctuations
  and non-Gaussianities in sidetracked inflation}},
  \href{https://doi.org/10.1088/1475-7516/2018/07/057}{\emph{JCAP} {\bfseries
  1807} (2018) 057} [\href{https://arxiv.org/abs/1804.11279}{{\ttfamily
  1804.11279}}].

\bibitem{Tolley:2009fg}
A.J.~Tolley and M.~Wyman, \emph{{The Gelaton Scenario: Equilateral
  non-Gaussianity from multi-field dynamics}},
  \href{https://doi.org/10.1103/PhysRevD.81.043502}{\emph{Phys. Rev.}
  {\bfseries D81} (2010) 043502}
  [\href{https://arxiv.org/abs/0910.1853}{{\ttfamily 0910.1853}}].

\bibitem{Dong:2010in}
X.~Dong, B.~Horn, E.~Silverstein and A.~Westphal, \emph{{Simple exercises to
  flatten your potential}},
  \href{https://doi.org/10.1103/PhysRevD.84.026011}{\emph{Phys. Rev. D}
  {\bfseries 84} (2011) 026011}
  [\href{https://arxiv.org/abs/1011.4521}{{\ttfamily 1011.4521}}].

\bibitem{McAllister:2014mpa}
L.~McAllister, E.~Silverstein, A.~Westphal and T.~Wrase, \emph{{The Powers of
  Monodromy}}, \href{https://doi.org/10.1007/JHEP09(2014)123}{\emph{JHEP}
  {\bfseries 09} (2014) 123} [\href{https://arxiv.org/abs/1405.3652}{{\ttfamily
  1405.3652}}].

\bibitem{Achucarro:2018vey}
A.~Ach{\'u}carro and G.A.~Palma, \emph{{The string swampland constraints
  require multi-field inflation}},
  \href{https://doi.org/10.1088/1475-7516/2019/02/041}{\emph{JCAP} {\bfseries
  1902} (2019) 041} [\href{https://arxiv.org/abs/1807.04390}{{\ttfamily
  1807.04390}}].

\bibitem{Cremonini:2010ua}
S.~Cremonini, Z.~Lalak and K.~Turzynski, \emph{{Strongly Coupled Perturbations
  in Two-Field Inflationary Models}},
  \href{https://doi.org/10.1088/1475-7516/2011/03/016}{\emph{JCAP} {\bfseries
  1103} (2011) 016} [\href{https://arxiv.org/abs/1010.3021}{{\ttfamily
  1010.3021}}].

\bibitem{Brown:2017osf}
A.R.~Brown, \emph{{Hyperbolic Inflation}},
  \href{https://doi.org/10.1103/PhysRevLett.121.251601}{\emph{Phys. Rev. Lett.}
  {\bfseries 121} (2018) 251601}
  [\href{https://arxiv.org/abs/1705.03023}{{\ttfamily 1705.03023}}].

\bibitem{Mizuno:2017idt}
S.~Mizuno and S.~Mukohyama, \emph{{Primordial perturbations from inflation with
  a hyperbolic field-space}},
  \href{https://doi.org/10.1103/PhysRevD.96.103533}{\emph{Phys. Rev.}
  {\bfseries D96} (2017) 103533}
  [\href{https://arxiv.org/abs/1707.05125}{{\ttfamily 1707.05125}}].

\bibitem{Achucarro:2017ing}
A.~Ach{\'u}carro, R.~Kallosh, A.~Linde, D.-G.~Wang and Y.~Welling,
  \emph{{Universality of multi-field $\alpha$-attractors}},
  \href{https://doi.org/10.1088/1475-7516/2018/04/028}{\emph{JCAP} {\bfseries
  1804} (2018) 028} [\href{https://arxiv.org/abs/1711.09478}{{\ttfamily
  1711.09478}}].

\bibitem{Christodoulidis:2018qdw}
P.~Christodoulidis, D.~Roest and E.I.~Sfakianakis, \emph{{Angular inflation in
  multi-field $\alpha$-attractors}},
  \href{https://doi.org/10.1088/1475-7516/2019/11/002}{\emph{JCAP} {\bfseries
  11} (2019) 002} [\href{https://arxiv.org/abs/1803.09841}{{\ttfamily
  1803.09841}}].

\bibitem{Linde:2018hmx}
A.~Linde, D.-G.~Wang, Y.~Welling, Y.~Yamada and A.~Ach{\'u}carro,
  \emph{{Hypernatural inflation}},
  \href{https://doi.org/10.1088/1475-7516/2018/07/035}{\emph{JCAP} {\bfseries
  1807} (2018) 035} [\href{https://arxiv.org/abs/1803.09911}{{\ttfamily
  1803.09911}}].

\bibitem{Garcia-Saenz:2018vqf}
S.~Garcia-Saenz and S.~Renaux-Petel, \emph{{Flattened non-Gaussianities from
  the effective field theory of inflation with imaginary speed of sound}},
  \href{https://doi.org/10.1088/1475-7516/2018/11/005}{\emph{JCAP} {\bfseries
  1811} (2018) 005} [\href{https://arxiv.org/abs/1805.12563}{{\ttfamily
  1805.12563}}].

\bibitem{Christodoulidis:2019mkj}
P.~Christodoulidis, D.~Roest and E.I.~Sfakianakis, \emph{{Attractors,
  Bifurcations and Curvature in Multi-field Inflation}},
  \href{https://doi.org/10.1088/1475-7516/2020/08/006}{\emph{JCAP} {\bfseries
  08} (2020) 006} [\href{https://arxiv.org/abs/1903.03513}{{\ttfamily
  1903.03513}}].

\bibitem{Bjorkmo:2019aev}
T.~Bjorkmo and M.C.D.~Marsh, \emph{{Hyperinflation generalised: from its
  attractor mechanism to its tension with the `swampland conditions'}},
  \href{https://doi.org/10.1007/JHEP04(2019)172}{\emph{JHEP} {\bfseries 04}
  (2019) 172} [\href{https://arxiv.org/abs/1901.08603}{{\ttfamily
  1901.08603}}].

\bibitem{Fumagalli:2019noh}
J.~Fumagalli, S.~Garcia-Saenz, L.~Pinol, S.~Renaux-Petel and J.~Ronayne,
  \emph{{Hyper-Non-Gaussianities in Inflation with Strongly Nongeodesic
  Motion}}, \href{https://doi.org/10.1103/PhysRevLett.123.201302}{\emph{Phys.
  Rev. Lett.} {\bfseries 123} (2019) 201302}
  [\href{https://arxiv.org/abs/1902.03221}{{\ttfamily 1902.03221}}].

\bibitem{Bjorkmo:2019fls}
T.~Bjorkmo, \emph{{The rapid-turn inflationary attractor}},
  \href{https://doi.org/10.1103/PhysRevLett.122.251301}{\emph{Phys. Rev. Lett.}
  {\bfseries 122} (2019) 251301}
  [\href{https://arxiv.org/abs/1902.10529}{{\ttfamily 1902.10529}}].

\bibitem{Christodoulidis:2019jsx}
P.~Christodoulidis, D.~Roest and E.I.~Sfakianakis, \emph{{Scaling attractors in
  multi-field inflation}},
  \href{https://doi.org/10.1088/1475-7516/2019/12/059}{\emph{JCAP} {\bfseries
  12} (2019) 059} [\href{https://arxiv.org/abs/1903.06116}{{\ttfamily
  1903.06116}}].

\bibitem{Bjorkmo:2019qno}
T.~Bjorkmo, R.Z.~Ferreira and M.D.~Marsh, \emph{{Mild Non-Gaussianities under
  Perturbative Control from Rapid-Turn Inflation Models}},
  \href{https://doi.org/10.1088/1475-7516/2019/12/036}{\emph{JCAP} {\bfseries
  12} (2019) 036} [\href{https://arxiv.org/abs/1908.11316}{{\ttfamily
  1908.11316}}].

\bibitem{Aragam:2019khr}
V.~Aragam, S.~Paban and R.~Rosati, \emph{{Multi-field Inflation in High-Slope
  Potentials}},
  \href{https://doi.org/10.1088/1475-7516/2020/04/022}{\emph{JCAP} {\bfseries
  04} (2020) 022} [\href{https://arxiv.org/abs/1905.07495}{{\ttfamily
  1905.07495}}].

\bibitem{Bravo:2019xdo}
R.~Bravo, G.A.~Palma and S.~Riquelme, \emph{{A Tip for Landscape Riders:
  Multi-Field Inflation Can Fulfill the Swampland Distance Conjecture}},
  \href{https://doi.org/10.1088/1475-7516/2020/02/004}{\emph{JCAP} {\bfseries
  02} (2020) 004} [\href{https://arxiv.org/abs/1906.05772}{{\ttfamily
  1906.05772}}].

\bibitem{Chakraborty:2019dfh}
D.~Chakraborty, R.~Chiovoloni, O.~Loaiza-Brito, G.~Niz and I.~Zavala,
  \emph{{Fat inflatons, large turns and the $\eta$-problem}},
  \href{https://doi.org/10.1088/1475-7516/2020/01/020}{\emph{JCAP} {\bfseries
  01} (2020) 020} [\href{https://arxiv.org/abs/1908.09797}{{\ttfamily
  1908.09797}}].

\bibitem{Aragam:2020uqi}
V.~Aragam, S.~Paban and R.~Rosati, \emph{{The Multi-Field, Rapid-Turn
  Inflationary Solution}},
  \href{https://doi.org/10.1007/JHEP03(2021)009}{\emph{JHEP} {\bfseries 03}
  (2021) 009} [\href{https://arxiv.org/abs/2010.15933}{{\ttfamily
  2010.15933}}].

\bibitem{Alishahiha:2004eh}
M.~Alishahiha, E.~Silverstein and D.~Tong, \emph{{DBI in the sky}},
  \href{https://doi.org/10.1103/PhysRevD.70.123505}{\emph{Phys. Rev.}
  {\bfseries D70} (2004) 123505}
  [\href{https://arxiv.org/abs/hep-th/0404084}{{\ttfamily hep-th/0404084}}].

\bibitem{Chen:2008hz}
X.~Chen, \emph{{Fine-Tuning in DBI Inflationary Mechanism}},
  \href{https://doi.org/10.1088/1475-7516/2008/12/009}{\emph{JCAP} {\bfseries
  0812} (2008) 009} [\href{https://arxiv.org/abs/0807.3191}{{\ttfamily
  0807.3191}}].

\bibitem{Ferreira:2020qkf}
R.Z.~Ferreira, \emph{{Non-Gaussianities in models of inflation with large and
  negative entropic masses}},
  \href{https://doi.org/10.1088/1475-7516/2020/08/034}{\emph{JCAP} {\bfseries
  08} (2020) 034} [\href{https://arxiv.org/abs/2003.13410}{{\ttfamily
  2003.13410}}].

\bibitem{Garcia-Saenz:2019njm}
S.~Garcia-Saenz, L.~Pinol and S.~Renaux-Petel, \emph{{Revisiting
  non-Gaussianity in multifield inflation with curved field space}},
  \href{https://doi.org/10.1007/JHEP01(2020)073}{\emph{JHEP} {\bfseries 01}
  (2020) 073} [\href{https://arxiv.org/abs/1907.10403}{{\ttfamily
  1907.10403}}].

\bibitem{Pinol:2020kvw}
L.~Pinol, \emph{{Multifield inflation beyond $N_\mathrm{field}=2$:
  non-Gaussianities and single-field effective theory}},
  \href{https://doi.org/10.1088/1475-7516/2021/04/002}{\emph{JCAP} {\bfseries
  04} (2021) 002} [\href{https://arxiv.org/abs/2011.05930}{{\ttfamily
  2011.05930}}].

\bibitem{Palma:2020ejf}
G.A.~Palma, S.~Sypsas and C.~Zenteno, \emph{{Seeding primordial black holes in
  multifield inflation}},
  \href{https://doi.org/10.1103/PhysRevLett.125.121301}{\emph{Phys. Rev. Lett.}
  {\bfseries 125} (2020) 121301}
  [\href{https://arxiv.org/abs/2004.06106}{{\ttfamily 2004.06106}}].

\bibitem{Fumagalli:2020adf}
J.~Fumagalli, S.~Renaux-Petel, J.W.~Ronayne and L.T.~Witkowski, \emph{{Turning
  in the landscape: a new mechanism for generating Primordial Black Holes}},
  \href{https://arxiv.org/abs/2004.08369}{{\ttfamily 2004.08369}}.

\bibitem{Fumagalli:2020nvq}
J.~Fumagalli, S.~Renaux-Petel and L.T.~Witkowski, \emph{{Oscillations in the
  stochastic gravitational wave background from sharp features and particle
  production during inflation}},
  \href{https://doi.org/10.1088/1475-7516/2021/08/030}{\emph{JCAP} {\bfseries
  08} (2021) 030} [\href{https://arxiv.org/abs/2012.02761}{{\ttfamily
  2012.02761}}].

\end{thebibliography}\endgroup

\end{document}